\documentclass[12pt]{article}
\input epsf.sty
\def\be{\begin{equation}} \def\ee{\end{equation}}
\def\bea{\begin{eqnarray}} \def\eea{\end{eqnarray}} \def\ba{\begin{array}}
\def\ea{\end{array}} \def\ben{\begin{enumerate}} \def\een{\end{enumerate}}

\newcommand{\eqn}[1]{(\ref{#1})}

\newcommand{\hepth}[1]{{\tt hep-th/{#1}}}

\def\l{\lambda}
\def\m{\mu}

\def\n{\nu}

\def\ov{\over}
\def\br{\nonumber\\}

\begin{document}
{}~
\hfill\vbox{\hbox{hep-th/yymm.nnnn} \hbox{\today}}\break

\vskip 2.5cm
\centerline{\large \bf
 $SU(N)$ membrane $B\wedge F$ theory  with dual-pairs }
\vskip .5cm

\vspace*{.5cm}

\centerline{  Harvendra Singh}

\vspace*{.25cm}
\centerline{ \it  Saha Institute of Nuclear Physics} 
\centerline{ \it  1/AF, Bidhannagar, Kolkata 700064, India}
\vspace*{.25cm}

\vspace*{.5cm}

\vskip.5cm
\centerline{E-mail: h.singh [AT] saha.ac.in }

\vskip1cm
%DRAFT : \today \\
\centerline{\bf Abstract} \bigskip

We construct a $SU(N)$ membrane $B\wedge F$ theory with  dual 
pairs of scalar and tensor fields. The moduli space of the theory 
 is precisely that of  
 $N$ M2-branes on the noncompact flat space.
The theory possesses  explicit $SO(8)$ invariance and is an extension of  
the maximal $SU(N)$ super-Yang-Mills theory.
  
\vfill \eject

\baselineskip=16.2pt

%%%%%%%%%%%%%%%%%%%%%%%%%%%%%%%%%%%%%%%%%%
%%%%%%%%%%%%%%%%%%%%%%%%%%%%%%%%%%%%%%%%%%%%%%

Interestingly,  recently 
certain type of matter Chern-Simons  field theories in 
$(1+2)$ dimensions have been 
proposed to be the low energy theories describing super-membranes. 
Amongst these, the 
originally proposed Bagger-Lambert-Gustavsson (BLG) theory  has ${\cal N}=8$ 
$SO(8)$ superconformal invariance but the theory is known only for $SO(4)$ 
tri-algebra  \cite{bl,Gustav}.  
Although for noncompact case of tri-algebras, BLG theory can be 
extended to admit  $SU(N)$ symmetry \cite{verlinde}. But these theories 
have  
ghost fields in the spectrum and once these are  gauge-fixed  the theory 
eventually reduces to the $SU(N)$  super Yang-mills 
\cite{Bandres:2008kj}.
 Another interesting class of matter-Chern-Simons theories proposed by 
Aharony-Bergman-Jafferis-Maldacena (ABJM) \cite{abjm}, however have  
ordinary
Lie-algebra structure. These theories admit 
${\cal N}=6$ 
$SU(N)_k\times SU(N)_{-k}$ superconformal 
symmetry, and is conjectured to be dual to  M-theory on $AdS_4\times 
S^7/Z_k$ with the level $k>2$. For $k=1,2$ the theory supposedly becomes 
maximally supersymmetric BLG theory.

It is now clear that the  understanding of Chern-Simons theories is 
essential
to know the M-theory origin of the $SU(N)$ Yang-Mills 
theory which describes $N$ D2-branes on $R^7$, and 
vice-versa. In particular, in the works \cite{bobby,mukhi} the authors 
have 
attempted to understand this link to some extent. The work \cite{bobby} 
is of 
particular 
importance to us in this paper. 
We take a parallel but 
rather distinct approach where we augment the B-F theory with 
scalars and dual 2-rank tensor fields, $ 
C_{(2)}^I$. This leads us to  a  membrane B-F  theory 
which has  $SU(N)$ gauge symmetry and has $SO(8)$ R-invariance as well as 
the scale invariance. The theory does not have any ghost degrees of 
freedom and also has no tri-algebras.     
It  presumably also has maximal supersymmetry as it is  
simply the topological 
extension of the 
$3D$ super Yang-Mills theory.

The low energy  $SU(N)$ super Yang-Mills 
theory with maximal supersymmetry is written as 
\bea
S_{SYM}&=&\int d^3x~ Tr (-{1\ov 4 g_{YM}^2} F_{\mu\nu}F^{\mu\nu} 
-{1\ov2}D_\mu 
X^iD^\mu X^i -{g_{YM}^2\ov 4} (X^{ij})^2 \br 
&&+{i\ov2}\bar\Psi 
\gamma^\mu D_\mu\Psi +{i\ov2}\bar\Psi\Gamma_i[X^i,\Psi])
\eea
where we defined $X^{ij}= [X^i,X^j]$. The field strengths are
\be
F_{\mu\nu}=\partial_{[\mu} A_{\nu]} -[A_\mu,A_\nu],~~~~~D_\mu 
X^i=\partial_\mu X^i-[A_\mu, X^i] \ee
The  bosonic fields $(A_\mu; X^1, ..., X^7)$  are all in the adjoint 
of $SU(N)$ and the fermions
$\Psi_\alpha^A$  form 2-compt spinor of $3D$ and 8-compt spinor of 
$SO(7)$. The theory has an explicit $SO(7)$ R-symmetry under which 
supercharges get rotated. The YM theory actually lives on the boundary 
of 
$AdS_4\times S^6$ (with varying string coupling), which is the near 
horizon geometry of $N$ D2-branes.
 
The scale (mass) dimensions are 
$$[X^i]={1\over 2},~[A_\mu]=1,~[\Psi]=1,~[g^2_{YM}]=1$$
Notice that the Yang-Mills coupling constant is dimensionful in $3D$ ! So 
the super Yang-Mills  can hardly be a conformal 
theory. In fact the YM coupling has a flow. Although the  theory  
has  good high energy 
behaviour where it becomes a free theory in UV regime, but in IR it is 
known 
to flow to a 
strongly coupled superconformal fixed point.
 The conformal nature of the YM theory at the IR fixed point has remained 
illusive though. Whether it describes  
 M2-brane theory has not been quite clear?  

For several reasons
it is expected that a theory of multiple M2-branes in flat space should 
have maximal supersymmetry,  should 
be conformal, should have $SO(8)$ R-symmetry and 
possibly  a gauge symmetry if it were an interacting theory. 
But the actual content of the theory has remained illusive so far.
A way ahead was suggested by the authors \cite{bobby} where one can 
make use of  
{\it de-Wit-Nicolai-Samtleben} duality transformations \cite{nicolai}. The 
dNS proposal is 
based on the fact that 
a propagating vector field in $3D$ contributes one  degree of freedom. 
It is a familiar kind of Poincare duality between gauge field  and a 
scalar field 
(${1\ov2!}\epsilon^{\mu\nu\lambda} F_{\nu\lambda}=\partial^\mu\phi$). 
Instead
in a  non-Abelian situation we can define
\be
{1\over 2! g_{YM}}\epsilon^{\mu\nu\lambda} 
F_{\nu\lambda}=D^\mu\phi-g_{YM} B^\mu
\ee
We can easily see that
\be
-{1\over4 g_{YM}^2}Tr (F_{\mu\nu})^2\equiv
 -{1\ov2}Tr(D^\mu\phi-g_{YM} B^\mu)^2 +Tr ({1\ov2}
\epsilon^{\mu\nu\lambda} 
B_\mu F_{\nu\lambda})
\ee
Thus the duality introduces a pair of adjoint fields $B_\mu$ and $\phi$. 
This duality has been used  in going from  $SO(7)$ 
 super-Yang-Mills  to the BLG theory which has $SO(8)$ R-symmetry 
\cite{bobby}. 
Actually after incorporating the dNS transformation the SYM Lagrangian 
takes the form
of a matter B-F (BF) Lagrangian
\bea
&&S_{BF}=\int d^3x \, Tr(
 -{1\ov2}(D^\mu\phi-g_{YM} B^\mu)^2 + {1\ov2}
\epsilon^{\mu\nu\lambda} B_\mu F_{\nu\lambda}
-{1\ov2}D_\mu 
X^iD^\mu X^i \br &&~~~~~~~-{g_{YM}^2\ov 4} (X^{ij})^2 +{i\ov2}\bar\Psi 
\gamma^\mu D_\mu\Psi +{i\ov2}\bar\Psi\Gamma_i[X^i,\Psi])
\eea
where we can now identify $\phi$ with $X^8$. This field alongwith the 
rest $X^i$'s forms an 
$SO(8)$ vector: $X^I~ (1\le I\le8)$. One then also defines a 
coupling constant 
8-vector:
$g^I=(0,\cdots,0,g_{YM})$.

With this we can write BF Lagrangian in an $SO(8)$ covariant Lagrangian 
form \cite{bobby}
\bea\label{cs}
&&S_{BF}=\int d^3x \, Tr(
 -{1\ov2}(D^\mu X^I-g^I B^\mu)^2 + {1\ov2}
\epsilon^{\mu\nu\lambda} B_\mu F_{\nu\lambda}-U(g^I,X^I)
\br &&~~~~~~~ +{i\ov2}\bar\Psi 
\gamma^\mu D_\mu\Psi +{i\ov2}g^I\bar\Psi\Gamma_{IJ}[X^J,\Psi])
\eea
where the potential is defined as 
\be
U={1\over 2.3!}V_{IJK}V^{IJK}
\ee
with the help of a completely antisymmetrized object
\be
V_{IJK}=g_{[I} X_{JK]}\equiv g_I X_{JK} + cyclic~ permutations 
\ee
Specially, we must note that parameters $g^I$ are in the $8_v$ while 
$X^{IJ}=[X^I,X^J]$ are 
in the 
adjoint of $SO(8)$ group. So as such the antisymmetrization of $V_{IJK}$
should not be confused with any tri-algebra like in BLG theory. However, 
it can be extended to have a Lorentzian tri-algebra structure 
\cite{bobby}.
 \footnote{Here 
the $SO(8)$ gamma matrices are 
$\Gamma_8=\tilde\Gamma^8,~\Gamma^i=\tilde\Gamma^8\tilde\Gamma^i$. (The 
matrices with tilde will henceforth will be named as $SO(7)$ matrices.)}

The action \eqn{cs} has an $SO(8)$ invariance provided the couplings 
$g^I$ transform along with  various fields under $SO(8)$ 
rotations. Thus,
although the theory has $SO(8)$ invariance but its action is transitive 
on the coupling parameters in the theory.  
After the transformations we get a new theory with a new set 
of couplings.
The ${\cal N}=8$ susy transformations can also be formally written in 
$SO(8)$ covariant form \cite{bobby}. 

It is noteworthy here to mention that such phenomena 
have also been observed  in the case of 
massive supergravity theory as well, see for an instance \cite{haack}.
In the present scenario,
the legitimate step would be like that in the Romans' theory in ten 
dimensions \cite{romans}.
There we try to lift the mass parameter (cosmological constant) $m$ to the 
level of a scalar field $M(x)$
which is then Hodge-dualised to a 10-form field strength $F_{10}$ 
\cite{berg}. This 
does not
introduce any new degree of freedom in the theory. Instead now the values 
of the mass parameters become localised in the spacetime. We shall like to 
implement
the same idea here for the $3D$ case. Note that we have  couplings $g^I$ 
in the vector 
representation of $SO(8)$. So we first define correspondingly 8 scalar 
fields 
$\eta^I(x)$ such that
\be
g^I=~ <\eta^I(x)>, ~~~~g^Ig^I=(g_{YM})^2
\ee
In the next step, we introduce 2-form potential $ C_{(2)}^I$, also 
in the $8_v$, whose 
field strength will 
be dual to $\eta^I$. We must also make sure that the  vacua are such that
 $\eta^I$ will be constant. This can be done simply by introducing a 
new topological term in the $SO(8)$ covariant BF action
\be
-\int  C_{(2)}^I \wedge d\eta^I
\ee
which is  $SO(8)$ 
invariant and  has the gauge invariance under
\be
  C_{(2)}^I \to  C_{(2)}^I + d\alpha_{(1)}^I \ee
Thus the complete membrane action can be written as
\footnote{ At this point we may be tempted to add another possible 
topological term
$-\theta\int C_{(3)}$, as it does not affect any of the dynamical 
considerations. Although from topological perspectives it  will be 
necessary.}~\footnote{ The $C_{(3)}$ can also be relevant while 
quantising 
the theory in the nontrivial membrane background. I thank S. Mukhi for 
this useful remark.}
   
\bea\label{MCS}
&&S_{MBF}=\int d^3x \, Tr(
 -{1\ov2}(D^\mu X^I-\eta^I B^\mu)^2 + {1\ov2}
\epsilon^{\mu\nu\lambda} B_\mu F_{\nu\lambda}-{1\over 2.3!}(V_{IJK})^2)
\br &&~~~~~~~~~~ -{1\ov2}\epsilon^{\mu\nu\lambda} C_{\mu\nu}^I 
\partial_\lambda\eta^I
+ S_{fermions} 
\eea
where 
$$D^\mu X^I=\partial^\m X^I - [A_\m,X^I], ~~~~~V_{IJK}=\eta_{[I} 
X_{JK]}\ .$$ The equations of motion are now 
augmented with two new set of equations. Namely the $ C^I$ equation
\be\label{new1}
\partial_\l\eta^I=0
\ee
and the $\eta^I$ equation
\be\label{new2}
 Tr((D^\mu X^I-\eta^I B^\mu)B_\mu - {1\over2} V^{IJK}X_{JK}) 
+{1\over 2}\epsilon^{\m\n\l} \partial_\m C_{\n\l}^I=0
\ee
The $ C^I$-equation implies that all $\eta^I$ are constant. The second 
equation only relates 
$\eta^I$ with its dual tensor field $ C_{\n\l}^I$ and should be taken as 
the duality relation. The rest of the 
field equations remain 
unchanged. So the net content of the  theory remains intact. There are 
no free parameters in the theory. The action 
\eqn{MCS} also has scale invariance. 
The 
supersymmetry presumably can also be made  manifest which we do not  
work out here. Henceforth we shall 
refer to the 
action \eqn{MCS} as membrane B-F (MBF) theory. 

Thus in bringing the  BF theory to the MBF form we have 
actually introduced dual 
pairs of fields $(C^I,\eta^I)$. The introduction of these dual pairs  has 
introduced a new paradigm in the MBF theory.
The moduli space of vacua in the MBF theory is now larger than the 
original SYM/BF theory. To know 
the moduli space of the MBF theory we  need to solve
\bea\label{new3}
&&\partial_{\eta^I} U(\eta,X)  -{1\ov2!}\epsilon^{\m\n\l} 
\partial_\m 
C_{\n\l}^I=0
\br&&~~~~~
 \rightarrow \eta^{[I} Tr( X^{JK]}X_{JK}) 
-\epsilon^{\m\n\l} \partial_\m C_{\n\l}^I=0
\eea
and
\be\label{eq16}
 \partial_{X^I} U(\eta,X)=0
\ee 
These equations have quite a few interesting possibilities. 

\noindent{\bf Case-1}: We take first $ C_{\m\n}^I=constt$. Since the 
solution 
of 
$\eta^I(x)=g^I$, we find that we 
need to have 
\be
X^{IJ}=[X^I,X^J]=0.
\ee
 This can happen when all the $X^I$'s are taken to be
diagonal matrices. That means all M2-branes are coincident.  
Hence the moduli space is exactly that of $N$ coincident M2-branes 
on noncompact $R^8$ space. 

However, the special case can arise when we take 
\be \eta^8=g_{YM},~\eta^i=0\ .
\ee
 This will then require 
\be X^{ij}=0~.
\ee 
In the simplest case 
all $X^i$ can be taken diagonal, but matrices $X^8$ can still be 
nontrivial but constant. These presumably will be the desired Goldstone 
modes corresponding to 
the spontaneously broken $SO(8)$ invariance. These will be  
eaten up by 
$B_\mu$ fields and making them heavy which can be integrated out 
in order to make the  $A_\mu$ 
fields 
dynamical. All this precisely corresponds to the moduli space of $N$ 
D2-branes on $R^7$. 

For both of the above  solutions the
components  $V_{IJK}$ are vanishing hence the scalar potential 
altogether vanishes. 
So these would make the maximally supersymmetric vacua in MBF theory. 

\noindent{\bf Case-2}: Another rather interesting case is of 3D 
domain-walls. Let 
us take the tensor components 
$ C_{01}^I$ to be  linearly dependent on one of the 
space coordinates, 
$x_2$ (say), 
then 
\be
d C^I\sim m^I dx^0\wedge dx^1\wedge dx^2
\ee
 is nontrivial, the 
$m^I$ being the slope parameters. The two 
such phases with different slopes can be separated 
via domain-walls which are just the line defects in 
2-dimensional plane. In
this situation, we shall have $g^I$ and $m^I$ related via   
\be\label{new4}
 {1\over2} g^{[I} Tr( X^{JK]}X_{JK}) -m^I=0
\ee
This will describe a noncommuting ({\it fuzzy}) configuration  of 
membranes. However, we are not sure if any static nontrivial fuzzy 
configuration can be found in 
which eq. \eqn{eq16} will be simultaneously satisfied. In any case, it 
will be interesting to find nonstatic solutions.

\noindent{\bf Quantisation:}

At this point, let us also discuss an interesting quantum aspect which 
follows straightforwardly from action \eqn{MCS}. The equation of motion 
for $X^I$, 
\be\label{eqxi}
{1\over 2! }\epsilon^{\mu\nu\lambda} 
F_{\nu\lambda}
=(D^\mu X^I-\eta^I B^\mu)\eta^I\ ,
\ee
and the equation \eqn{new2} can be combined to give
\be\label{eqquan}
 Tr({1\over 2! }\epsilon^{\mu\nu\lambda} 
F_{\nu\lambda}B_\mu - U) 
+{1\over 2}\eta^I\epsilon^{\m\n\l} \partial_\m C_{\n\l}^I=0
\ee
In the vacuum where $U=0$, it has interesting implications. 
For 
example, consider an Euclidean monopole configuration where $F_{\m\n}\ne 
0$  inside a 
3-dimensional volume $V^3$, with a boundary $\partial V^3\sim S^2$. 
We can have a configuration where
\be
Tr{1\over 4\pi }\int_{V^3} B\wedge F \sim {p(N)} \ , \ee 
Here we have taken  $p(N)\in {\bf Z}$ 
to depend upon the rank $N$ of the 
Yang-Mills group. The actual expression of $p(N)$ however will depend 
upon the details of 
the monopole configuration.
We are taking $SO(7)$ configuration where 
$\eta^8=g_{YM},~\eta^i=0$.
The  equation \eqn{eqquan} leads us to 
the 
quantization 
\be
-{1\over 4\pi \sqrt{l_p}}\int_{V^3} dC_{(2)}
=-{1\over 4\pi \sqrt{l_p}}\int_{S^2} C_{(2)}
= k \in {\bf Z}\ .
\ee
with $g_{YM}\sim {p(N)\over(l_p)^{1/2} k}$, and we introduced $l_p$, 
the 11-dimensional Planck length. 
 That is  we need to 
have a nontrivial $ C_{(2)}$ flux over $S^2$.
It does mean that Yang-Mills coupling 
in a given topological vacuum is controlled by the ratio of $p(N)$ and 
$k$. By
having large  $k$ limit we can accomodate a weak Yang-Mills coupling.  
This argument appears almost analogous to  large $k$ limit in $C_4/Z_k$ 
orbifold models 
\cite{abjm}.

In summary, we have taken an approach
where we augmented the B-F theory with 
scalars and dual 2-rank tensor fields, namely $(\eta^I,~ 
C_{(2)}^I)$. This led us to  a  membrane B-F  theory 
which has  $SU(N)$ gauge symmetry and has $SO(8)$ R-invariance as well as 
the scale invariance. There are no free parameters in the action. The 
theory does not have any ghost degrees of 
freedom and also has no tri-algebras. So in that aspect our theory is 
distinct from the 
B-F theory of \cite{bobby}.     
Our theory presumably also has maximal supersymmetry as it is  
simply the topological extension of the 
$3D$ super Yang-Mills theory. Interestingly, the moduli space comes out to 
be that of $N$ 
coincident M2-branes on transverse $R^8$.

\centerline{---------------------}

\noindent{\it Acknowledgements:}\\
I am very much thankful to S. Roy and specially S. Mukhi for several 
helpful 
discussions.

%\newpage


\begin{thebibliography}{99}
\bibitem{bl} J. Bagger and N. Lambert, \hepth{0611108}; \hepth{0711.0955}; 
\hepth{0712.3738}.

\bibitem{Gustav}
  A.~Gustavsson,
  %``Algebraic structures on parallel M2-branes,''
  arXiv:0709.1260 [hep-th]; A.~Gustavsson,
  %``Selfdual strings and loop space Nahm equations,''
  JHEP {\bf 0804}, 083 (2008)
  [arXiv:0802.3456 [hep-th]].

\bibitem{verlinde}
  J Gomis, G. Milanesi and J.G. Russo, arXiv:0804.1012; S.~Benvenuti, 
D.~Rodriguez-Gomez, E.~Tonni and H.~Verlinde,
  %``N=8 superconformal gauge theories and M2 branes,''
  arXiv:0805.1087 [hep-th].
J. Gomis, D. Rodriguez-Gomez, M. Van Raamsdonk, H. Verlinde,
arXiv:0806.0738v2 [hep-th]


\bibitem{Bandres:2008kj}
  M.~A.~Bandres, A.~E.~Lipstein and J.~H.~Schwarz,
  %``Ghost-Free Superconformal Action for Multiple M2-Branes,''
  arXiv:0806.0054 [hep-th];
  M.~A.~Bandres, A.~E.~Lipstein and J.~H.~Schwarz,
  %``N = 8 Superconformal Chern--Simons Theories,''
  JHEP {\bf 0805}, 025 (2008)
  [arXiv:0803.3242 [hep-th]].

\bibitem{abjm}
  O.~Aharony, O.~Bergman, D.~L.~Jafferis and J.~Maldacena,
  ``N=6 superconformal Chern-Simons-matter theories, M2-branes and their
  gravity duals,''
  arXiv:0806.1218 [hep-th].

\bibitem{bobby}
B.~Ezhuthachan, S.~Mukhi and C.~Papageorgakis,
  %``D2 to D2,''
  JHEP {\bf 0807}, 041 (2008)
  [arXiv:0806.1639 [hep-th]].

\bibitem{mukhi}
  S.~Mukhi and C.~Papageorgakis,
  %``M2 to D2,''
  JHEP {\bf 0805}, 085 (2008)
  [arXiv:0803.3218 [hep-th]];  
J.~Distler, S.~Mukhi, C.~Papageorgakis and M.~Van Raamsdonk,
  %``M2-branes on M-folds,''
  JHEP {\bf 0805}, 038 (2008)
  [arXiv:0804.1256 [hep-th]].


\bibitem{nicolai} H. Nicolai and H. Samtleben, Nucl.Phys. B668 (2003) 
167-178, [arXiv:0303213 [hep-th]];
B. de Wit, I. Herger, H. Samtleben,
Nucl.Phys. B671 (2003) 175-216
arXiv:hep-th/0307006;
B. de Wit, H. Nicolai, H. Samtleben,
 arXiv:hep-th/0403014.
\bibitem{haack}
  M.~Haack, J.~Louis and H.~Singh,
  %``Massive type IIA theory on K3,''
  JHEP {\bf 0104}, 040 (2001)
  [arXiv:hep-th/0102110];
H.~Singh,
  %``Duality symmetric massive type II theories in D = 8 and D = 6,''
  JHEP {\bf 0204}, 017 (2002)
  [arXiv:hep-th/0109147].
\bibitem{romans} L. J. Romans, Phys. Lett. B169 (1986) 374.
\bibitem{berg} E. Bergshoeff, M. de Roo, M. Green, G. Papadopoulos and P. 
Townsend,
Nucl. Phys. B470 (1996) 113, arXiv:hep-th/9601150 .


\end{thebibliography}
\end{document}